\documentclass[12pt,preprint]{aastex}

\begin{document}

\title{Low-luminosity Active Galaxies and their Central Black Holes}
\author{X.Y. Dong and M.M. De Robertis}
\affil{Department of Physics and Astronomy, York University, 4700
Keele St., Toronto, ON, M3J 1P3} \email{xydong@yorku.ca,
mmdr@yorku.ca}

\begin{abstract}
Central black hole masses for 118 spiral galaxies representing
morphological stages S0/a through Sc and taken from the large
spectroscopic survey of \citeauthor{Ho1997} \citeyearpar{Ho1997} are
derived using $K_s$-band data from 2MASS.  Black hole masses are
found using a calibrated black-hole -- $K_s$ bulge luminosity
relation, while bulge luminosities are measured by means of a
two-dimensional bulge/disk decomposition routine.

The black hole masses are correlated against a variety of parameters
representing properties of the nucleus and host galaxy. Nuclear
properties such as line width (FWHM([N\,{\small II}])), as well as
emission-line ratios (e.g., [O\,{\small III}]/H$\beta$, [O\,{\small
I}]/H$\alpha$, [N\,{\small II}]/H$\alpha$, and [S\,{\small
II}]/H$\alpha$), show a very high degree of correlation with
black-hole mass.  The excellent correlation with line-width supports
the view that the emission-line gas is in virial equilibrium with
either the black hole or bulge potential. The very good
emission-line ratio correlations may indicate a change in ionizing
continuum shape with black hole mass in the sense that more massive
black holes generate harder spectra.

Apart from the inclination-corrected rotational
velocity, no excellent correlations are found between black-hole
mass and host-galaxy properties.

Significant differences are found between the distributions of black
hole masses in early-, mid- and later-type spiral galaxies
(subsamples A, B and C) in the sense that early-type galaxies have
preferentially larger central black holes, consistent with
observations that Seyfert galaxies are found preferentially in
early-type systems.  The line-width distributions show a marked
difference among subsamples A, B and C, in the sense that
earlier-type galaxies have larger line widths.  There are also clear
differences in line ratios between subsamples A+B and C that likely
are related to the level of ionization in the gas.  Finally, a
$K_s$-band Simien \& de Vaucouleurs diagram shows excellent
agreement with the original $B$-band relation, though there is a
large dispersion at a given morphological stage.

\end{abstract}

\keywords{galaxies: bulges --- galaxies: fundamental parameters ---
infrared: galaxies --- galaxies: photometry --- galaxies: active}

\section{Introduction}

Activity in galactic nuclei is almost certainly the result of the
accretion of gas onto supermassive black holes \citep{Rees1984}.
While the bolometric luminosity of some active galactic nuclei
(AGNs) can approach and even exceed the luminosity of their host
galaxies in the case of the brightest Seyfert galaxies and QSOs, it
has been recognized for some time that there are AGNs of
considerably lower luminosity.  A large-scale spectroscopic survey
by \citeauthor{Ho1997} \citeyearpar{Ho1997} (hereafter, HFS), for
example, discovered that approximately 86\% of all galaxies brighter
than an apparent $B$ magnitude of +12.5 contain detectable
emission-line nuclei, and 43\% of all galaxies that fall in its
survey limits can be considered `active'. For galaxies with an
obvious bulge component, this rises from 50\% to 70\% for early type
galaxies, i.e., ellipticals, lenticulars, and bulge-dominated
spirals (S0/a-Sbc).

There is now excellent evidence that supermassive black holes (BH)
are found in the centers of dozens of nearby galaxies
\citep{Kormendy2004}, including our own \citep{Ghez2004}. Moreover,
there is a rather tight correlation between the mass of the BH and
the velocity dispersion of the spheroidal stellar component in which
it is situated (e.g., \citeauthor{Ferrarese2000}
\citeyear{Ferrarese2000},
\citeauthor{Gebhardt2000}\citeyear{Gebhardt2000},
\citeauthor{Gebhardt2003}\citeyear{Gebhardt2003}) and a good
correlation between the mass of the bulge and BH. Under the
assumption that this relation extends to all relatively bright,
nearby galaxies, it is possible to probe the relationship between
BHs and the host galaxies harboring low-luminosity AGNs.  By
studying this population in some detail, it is hoped that a greater
insight will be achieved into how nuclear activity originates and is
maintained.

In this paper, we use near-infrared $K_s$-band imaging data from the
Two Micron All Sky Survey (2MASS)\footnote[1]{\url{http://www.ipac.caltech.edu/2mass}} for 118 spiral
galaxies (from S0/a to Sc) in the HFS sample to measure bulge
luminosities and hence derive BH masses.  There are definite
advantages to working in the near infrared for our purposes:
compared with the optical band, extinction effects in the $K$-band
are considerably reduced \citep[e.g.,][]{Schlegel1998}, while the
composite spectral energy distribution of bulge stars is well
represented in the $K$-band \citep[e.g.,][]{Bendo2004}. Correlations
are sought between the BH mass and a number of properties of the
galaxy, including structural properties, as well as properties of
the nucleus, including `active' and `non-active' parameters.
Comparisons of the distributions of the various parameters as a
function of morphological type are also explored.

Section 2 describes the dataset which forms the basis of this study.
Section 3 provides details on the bulge-disk decomposition of the
2MASS data, as well as the calibration of the $K_s$-band bulge-BH
relation.  In section 4, we discuss the good-to-excellent
correlations between BH mass and various galaxy and nuclear
parameters as well as comparisons of the various distributions,
while we offer opinions on the reasons for these correlations in
section 5, the summary and conclusion.

\section{The HFS Palomar Survey}

We selected 118 galaxies from among the 486 nearby galaxies observed
in the high-quality HFS spectroscopic survey carried out with the 5
m telescope at Mt. Palomar. A primary goal of the HFS study was to
uncover `dwarf Seyfert nuclei' and to determine their luminosity
function. The authors selected galaxies from the Revised
Shapley-Ames Catalogue of Bright Galaxies
\citep[RSA;][]{Sandage1981} and the Second Reference Catalogue of
Bright Galaxies \citep[RC2;][]{deVaucouleurs1976} subject to the
criteria that the total apparent $B$ magnitude $B_T \leq +12.5$ and
declination $\delta
> 0^{\circ}$.  Incompleteness of the RSA catalog sets in near $B_T
= +12$ mag and becomes increasingly more severe by $B_T = +12.5$
mag. The Palomar AGNs have a median (narrow) $\rm{H\alpha}$
luminosity of $2\times10^{39}$ erg~s$^{-1}$ (HFS). Typical Seyfert
nuclei in the Markarian catalog emit $\sim 10^{41}$ erg s$^{-1}$ in
$\rm{H\alpha}$ (\citeauthor{Barth1999} \citeyear{Barth1999}).
Galaxies with $\rm{ L(H\alpha)}\leq10^{40}$ erg s$^{-1}$ are
considered `low-luminosity' or `dwarf' AGNs.

\subsection{Our sample}

Galaxies in our sample were selected from HFS and divided into three
subsamples according to morphology or Hubble stage parameter, $T$.
Subsample A includes only early-type spirals, S0/a and Sa galaxies,
with $T = 0 - 1$. Subsample B includes intermediate-type spirals,
Sab, Sb, Sbc galaxies, with $T = 2 - 4$, while subsample C includes
late-type spirals, Sc galaxies, with $T = 5$.

Galaxies which do not have photometry from 2MASS are not included in
our sample.  Galaxies in the Virgo cluster are excluded if their
projected angular separations to either of the two Virgo centers
(i.e., M87 and NGC 4472) is $< 3^\circ$, in order to reduce possible
uncertainties resulting from environmental effects.  (Only four 
galaxies were rejected based on this criterion.) 
Sample C galaxies that appear only to have a disk
component are excluded as well.  (About 15\% of the galaxies of the
appropriate morphological type were rejected from Sample C 
because they did not have a perceptible bulge component.)
In our final sample, subsample A contains 38 galaxies, while
subsamples B and C contain the first 40 galaxies from the HFS
catalog satisfying the other selection criteria.

The data (Table \ref{em-host}) for the survey galaxies taken from
HFS include emission-line parameters and host galaxy parameters.
Distances were adopted from
\citeauthor{Tully1988}\citeyearpar{Tully1988} and are derived from a
Virgo infall model \citep{Tully1984} based on $\rm H_o = 75$ km
s$^{-1}$ Mpc$^{-1}$, the value of the Hubble constant used
throughout this paper. Refer to HFS for details concerning the
measurement and reduction techniques.

\subsection{2MASS data}

The Two Micron All Sky Survey (2MASS) consists of exposures for 98\%
of the sky simultaneously in $J$ (1.24 $\rm{\mu m}$), $H$ (1.66
$\rm{\mu m}$), and $K_s$ (2.16 $\rm{\mu m}$). The total integration
time of each image is 7.8 s with point-source sensitivity limits (10
$\sigma$) of 15.8 (0.8 mJy), 15.1 (1.0 mJy), and 14.3 (1.4 mJy) mag
at $J$, $H$, and $K_s$, respectively. The extended source
sensitivity limits (10 $\sigma$) are 14.7 (2.1 mJy), 13.9 (3.0 mJy),
and 13.4 (4.1 mJy) mag at $J$, $H$, $K_s$, respectively.

Total magnitudes for all of our galaxies were taken from the
Extended Source Catalog along with galaxy positions, photometry and
basic shape information.  The All Sky Extended Source Image Server
provides full-resolution images in the angular size ranging from
$21^{\prime\prime} \times 21^{\prime\prime}$ to $301^{\prime\prime}
\times 301^{\prime\prime}$. Galaxies larger than $2^\prime$ or
$3^\prime$ are collected in the 2MASS Large Galaxy Atlas. The plate
scale of the 2MASS data is 1 arcsec pixel$^{-1}$. For our 118 sample
galaxies, 69 were taken from the All Sky Extended Source Image
Server and 49 from the Large Galaxy Atlas.

\subsection{Measurement of Black Hole Mass}

A variety of methods have been used to measure black-hole masses in
galactic nuclei.  Primary methods employ either stellar or gas
kinematics to derive dynamical masses.  Stellar dynamical techniques
are used because stars are always present and their motions are
subject to the gravitational potential. But they can be used only
for nearby galaxies since high spatial resolution is required to
measure the velocity dispersion of stars around the center of the
galaxy \citep[e.g.,][]{Kormendy2004}. These techniques are not
suitable for bright ellipticals and when a dust disk is present. 
Biases and systematics of the stellar kinematics technique 
can prove severe as well
\citep[e.g.,][]{Valluri2004}.  Gas dynamics from water maser clouds
can provide good BH mass measurements \citep[e.g.,][]{Greenhill1995,
Moran1999}; unfortunately, $\rm{H_{2}O}$ masers are not sufficiently
common. Gas dynamics of nuclear dust/gas disks can be used in early
type galaxies where stellar dynamical studies fail.  But as
\citeauthor{Kormendy1995}\citeyearpar{Kormendy1995} have noted,
gas, unlike stars, will respond to non-gravitational forces, and the
motions of gas clouds may not always reflect the underlying
gravitational potential.  Another primary method for the measurement
of black-hole masses is reverberation mapping
\citep[e.g.,][]{Blandford1982, Wandel1999, Kaspi2000}.

Secondary methods include scaling relationships based on
reverberation methods \citep[e.g.,][]{Peterson2004}.

\subsection{BH Mass vs.~Bulge Correlation}

The mass of the central BH correlates well with the $B$-band
luminosity of the bulge component of its host galaxy, and it
correlates even more strongly with the velocity dispersion of the
bulge. \citeauthor{Tremaine2002} \citeyearpar{Tremaine2002} and
\citeauthor{Gebhardt2003}\citeyearpar{Gebhardt2003} demonstrated a
strong correlation between the mass of the BH and the velocity
dispersion of the host galaxy using high quality data from 31 nearby
galaxies selected from
\citeauthor{Ferrarese2000}\citeyearpar{Ferrarese2000},
\citeauthor{Gebhardt2000} \citeyearpar{Gebhardt2000},
\citeauthor{Merritt2001}\citeyearpar{Merritt2001}, and
\citeauthor{Kormendy2001}\citeyearpar{Kormendy2001}.

\begin{figure}
\begin{center}
\includegraphics[scale=.8]{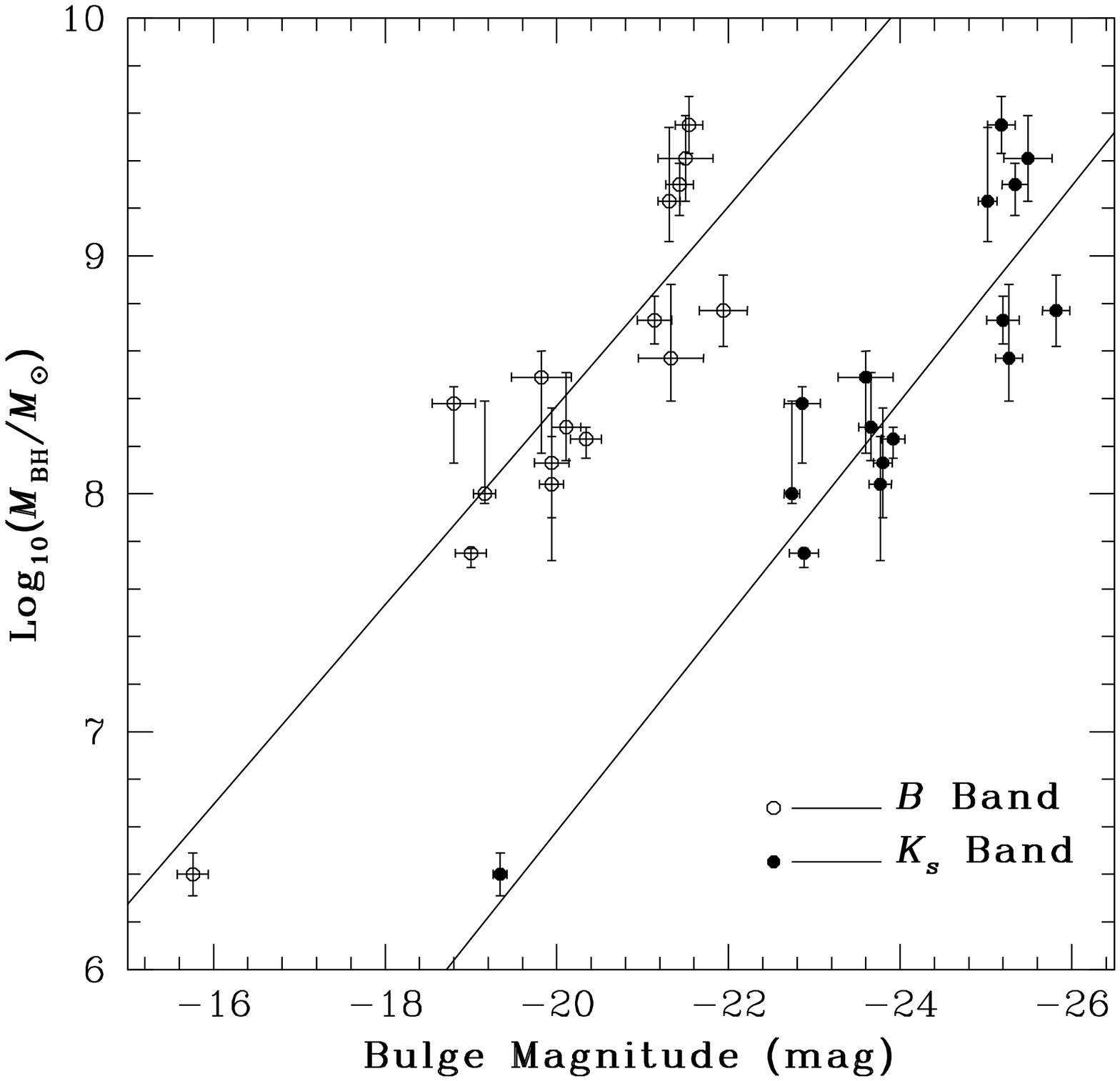}
\caption{BH mass vs.~the bulge magnitude for the $B$-band (open
dots) and $K_{\rm s}$-band (filled dots). Solid lines indicate the
linear least-squares fitting results. Uncertainties in magnitude and
mass are illustrated. \label{BH-B-K}}
\end{center}
\end{figure}

We performed a weighted least-squares $K_s$-BH mass fit using 2MASS
$K_s$ magnitudes for 16 elliptical galaxies whose black hole masses
(with uncertainties) and distances were taken from the compilation
of \citeauthor{Ferrarese2005} \citeyearpar{Ferrarese2005} and reproduced
in Table \ref{BH-calibration}:

\begin{equation}\label{Ks-BH}{\log_{10}(M_{\rm {BH}}/M_{\rm \odot}) = (-0.45\pm0.03)\, K_{s} + (-2.5\pm0.6)}\end{equation}
(Three elliptical galaxies from \citeauthor{Ferrarese2005}
\citeyearpar{Ferrarese2005} were not included in the fit: NGC 2778
has an S0 light profile, the radius of the sphere of influence for 
NGC 821 is much smaller than the spatial resolution, and Cyg A is
in a very rich cluster environment.) 
Compared with the $B$-band fit provided in
\citeauthor{Ferrarese2005}\citeyearpar{Ferrarese2005}, the
$K_s$-band relation has only a slightly smaller dispersion. The
linear correlations in the $K_s$-band and the $B$-band fits are
illustrated in Figure \ref{BH-B-K}.

Equation (1) is derived under the assumption that elliptical
galaxies behave as pure spheroids; i.e., scale as spiral galaxy
bulges and so the total magnitudes from the XSC can be used
directly. The absolute magnitudes are calculated from the apparent
magnitudes using the distances provided in \citeauthor{Tonry2001}
\citeyearpar{Tonry2001} with the appropriate Hubble constant change.
Our relation is consistent with that derived by
\citeauthor{Marconi2003}\citeyearpar{Marconi2003} who used both
elliptical and spiral galaxies.  We chose not to include spiral
galaxies in our calibration, 
preferring to minimize uncertainties introduced by 
bulge-disk decomposition.

\section{Bulge-Disk Decomposition}\label{B-D-decomposition}

It is necessary to decompose the bulge and disk luminosities before
using the bulge luminosity-BH mass relation to estimate the BH mass
for each of our sample spiral galaxies.  We chose to employ a
two-dimensional bulge-disk decomposition technique on the 2MASS
$K_s$-band images.  There is evidence that this technique is more
robust than one-dimensional surface-brightness profile fitting,
particularly when considering the uniqueness of a fit, and the
effects of asymmetric light distributions.

We used GALFIT (\citeauthor{Peng2002} \citeyear{Peng2002}) to
decompose galaxies into a spheroidal bulge following a S\'{e}rsic
profile \citep{Sersic1968} and an exponential disk
\citep{Freeman1970}. (GALFIT can accommodate as many components as
required to generate a cleaner residual image.  Unless there is a
compelling physical reason to add another component, however, it is
best to include only these two components.)  A model galaxy image is
created based on initial input parameters that can be convolved with
a Point Spread Function (PSF) image before comparison with the
actual galaxy image. Fitting proceeds iteratively until convergence
is achieved, which normally occurs when $\chi^2$ does not change by
more than 5 parts in $10^4$ for five iterations. (See \citeauthor{Peng2002} \citeyearpar{Peng2002} for the discussion of
the profile functions and fitting parameters.)

Only the S\'{e}rsic profile and the exponential profile were used
for a galaxy fit unless there was clear evidence from the residual
image for the existence of another component.  Fewer than 16\% of
galaxies in our sample showed a more complicated structure. A mask
was used in the centers of the 18 Seyfert galaxies in our sample to
avoid possible contamination from the nuclear point source,
something that will be discussed later.

\subsection{Using GALFIT}

Initial values for many of the parameters used in the fitting
process could be taken from the 2MASS FITS image header, including
the galaxy's X and Y centers and zero point magnitude. The total
magnitude of the galaxy was taken from the 2MASS XSC. The initial
value for the bulge magnitude was estimated from the empirical
relationship \citep{Simien1986}:
\begin{equation}\label{mbul} m_{bul}=m_{tot}+\Delta m_{bul}\end{equation}
and
\begin{equation}m_{disk}=m_{bul}+2.5log_{10}\Gamma\end{equation}
where
\begin{equation}\label{dmbul} \Delta m_{bul}=0.324(T+5)-0.054(T+5)^2+0.0047(T+5)^3\end{equation}
is the magnitude difference between the spheroid component and the
total galaxy, $T$ is the Hubble morphological parameter. $\Gamma$ is
the bugle-to-disk luminosity ratio, $\Gamma\equiv\kappa/(1-\kappa)$,
where $\kappa$ is the bulge-to-total luminosity ratio, and
$\kappa=10^{-0.4\Delta m_{bul}}$.

The PSF images were generated from 2MASS images using tasks in the
DAO Crowded-Field Photometry Package (DAOPHOT) of IRAF\footnote[2]{IRAF 
is distributed by the National Optical Astronomy
Observatories, which are operated by the Association of Universities
for Research in Astronomy, Inc., under cooperative agreement with
the National Science Foundation.}. A convolution radius of 20 seeing
disks was used (Peng, private communication).

The initial value for $n$, the S\'{e}rsic index, was taken to be 4,
i.e., a de Vaucouleurs bulge.  The axial ratio $q$ was taken from
the 2MASS XSC, fitted to the $3\sigma$ isophote.  The position angle
was also taken from the 2MASS XSC.  The shape parameter, $c$, was
fixed at 0.

Initial estimates for the bulge and disk scale radii, $R_e$ and
$R_s$ ($R_e=1.678 R_s$), were taken from the 2MASS XSC. $R_e$ and
$R_s$ are sensitive to the background level. An ill-defined
background level can cause an unreasonably large or small $R_s$,
which will affect the bulge parameters as well.  2MASS XSC images
are individually background-subtracted using a weighted cubic
polynomial smoothing technique on angular scales larger than our images \cite{Jarrett2000}.  Unable
to provide a better background estimate, we set the sky value to 0
during the fitting process. The vast majority of galaxies had $R_e$
larger than the effective seeing disk; $2.5^{\prime\prime}$. Fewer
than 3\% had $R_e$ less than one seeing disk. Since the S\'{e}rsic
index `n' and $R_e$ are coupled
--- small $n$ always has small $R_e$ \cite{Graham2001} --- we
accepted the best-fit $R_e$ that is larger than 0.5 kpc
\cite{Graham2001} and less than $R_s$.

The 4 Seyfert 1s and 14 Seyfert 2s galaxies in our sample are the
only objects whose derived bulge magnitudes could possibly have been
compromised significantly by a central component (though no obvious
point source was observed in the spatial profiles of these
galaxies).  Without knowing a priori the relative contribution of
the point source, it seemed prudent to insert a circular mask of
radius $3^{\prime\prime}$ (slightly larger than the FWHM of the PSF)
at the center of each Seyfert galaxy. Only data outside of this mask
were used in the fit. Some empirical experiments showed that while
masking might lead to systematically fainter bulges given the
relatively low spatial resolution of 2MASS data, the average
difference is within the uncertainty reported by GALFIT. Two tests
were also performed to investigate how $m_{bul}$ changes if the fit
is perturbed somewhat from the best fit by varying the S\'{e}rsic
index $n$.  In the majority of cases, varying $n$ by factors of two
leads to bulge luminosity differences less than 10\%. The bulge
luminosity is then fairly robust.

GALFIT's output includes uncertainties for each parameter in an
output file.  The uncertainties provided by GALFIT are equal to or
slightly larger than the uncertainties determined through extensive
modeling (\citeauthor{Peng2002} \citeyear{Peng2002}).  There clearly
are limitations in determining parameter errors, and uniqueness is
never guaranteed. A good fit occurs, very frequently, when $\chi^2$
is minimized.  But since $\chi^2$ depends on the number of degrees
of freedom of the fit, it is difficult to decide upon an optimal fit
based solely on $\chi^2$, especially when some parameters are fixed
or constrained during the fitting.  Other criteria that were used to
assess the quality of the fit include assessing: a) the physical
reasonability of the parameter(s), b) the smoothness of the residual
image, c) the surface brightness profiles to see if the sum of the
components fits the actual galaxy well, and d) the differences
between the total magnitude of the models and actual galaxies to
ensure they are small and not systematic.  For our sample, 63\% of
galaxies have magnitude differences within $\pm$0.1, while fewer
than 11\% have differences larger than $\pm$0.2.

Bulge-disk decomposition can be problematic in edge-on systems for a
variety of reasons. Fortunately, only 6\% of our sample galaxies
have inclination angles greater than 75$^\circ$.

The GALFIT results are given in Table \ref{GALFIT-result}. From the
GALFIT results, the total brightness of the bulge and the disk are
converted to absolute magnitudes, while the scale lengths are
converted to kiloparsecs using distances provided in HFS. From the
absolute bulge magnitude, the BH mass can be derived.  Table
\ref{BH-mass} gives the final results including the BH mass, the
absolute magnitudes of the bulge, disk and the total galaxy, and the
structure parameters (i.e., the scale lengths) of each component of
the galaxy.

Uncertainties of the estimated BH masses include the uncertainty in
the slope of the logarithm of the BH mass - bulge correlation, the
formal uncertainties in the decomposition of the bulges, and the
distance uncertainties (which provide the largest contribution).
When considered together, the uncertainty in the logarithm of the
black hole mass is $\pm$0.27 or about a factor of 2.

\section{Discussion}

In this section we describe the correlations found in this dataset
and discuss their possible physical significance.

\subsection{Correlations}

A linear least-squares technique was used to search for correlations
between parameters.  In particular, we assessed the significance
level or probability, $P$, that a linear correlation coefficient of
the appropriate magnitude could happen by chance from a random
distribution.

The most significant correlations are provided in Table
\ref{correlations}. An excellent correlation has $P<10^{-4}$, a good
correlation has $10^{-4}<P<0.05$, while a poor correlation has $P\ge
0.05$.

It should be noted that quantities recorded in HSF as either upper
or lower limits were not used when computing any correlation
coefficient.  Since the number of values with upper or lower limits
are fewer than 10\% for any parameter set considered herein,
excluding galaxies with upper or lower limits is not expected to
change our results significantly.

The nuclear properties, i.e., the emission-line ratios and the
FWHM([N\,{\small II}]), are all well correlated with the BH mass,
but they are not correlated with host galaxy properties except for
the bulge luminosity which was anticipated.

\begin{figure}
\centering
\includegraphics[scale=.8]{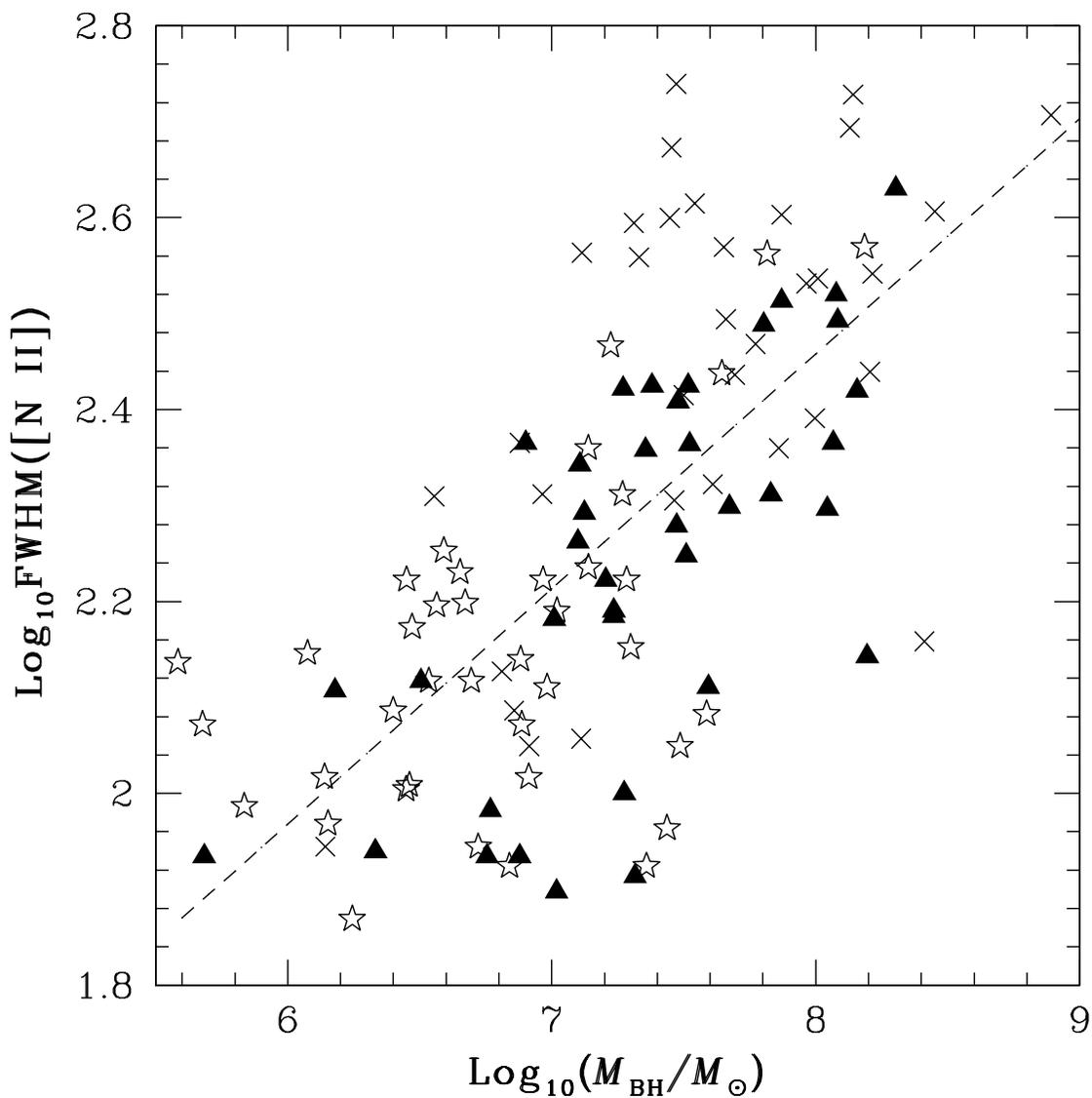}
\caption{BH mass vs.~FWHM([N\,{\small II}]) for sample A (crosses),
sample B (filled triangles), and sample C (open asterisks) galaxies.
The dashed line is the linear least-squares result considering all
three samples together. \label{BH-FWHM}}
\end{figure}

The excellent correlation between emission-line width and BH mass
($P<10^{-4}$) may provide support for the model in which the
emission lines are generated by gas in the nucleus in virial
equilibrium as has been hypothesized for more luminous systems
\citep[e.g.,][]{Nelson1996, Boroson2003}. If gas clouds are in
virial equilibrium within a radius $R$, then the three-dimensional
dispersion velocity $\rm{\sigma}$ follows:
\begin{equation}{\sigma^2 = G \it M(R)/R}\end{equation}
where $M(R)$ is the mass within the radius $R$, and $G$ is the
gravitational constant. Figure \ref{BH-FWHM} shows the correlation
between the BH mass and the FWHM([N\,{\small II}]). The FWHM of an
emission line is related to the three-dimensional velocity
dispersion FWHM$ = 2.35\times\sigma/\sqrt{3}$. It is not clear from
these data, however, whether the BH or the bulge dominates the
gravitational field in which the emission-line material moves.  If
the gravitational field is dominated by a BH with a mass about
$10^7$ solar masses, a line width of 200 km~s$^{-1}$ will result
from gas clouds at $R_{\rm BH} \sim 0.6$ pc. If the gravitational
field is dominated by a $\sim 10^{10}$ solar mass bulge of radius
2~kpc, a FWHM of 200 km~s$^{-1}$ arises from gas at $\sim$ 1 kpc.

Adequate spatial resolution is required to determine which
component, the BH or the bulge, dominates the gas-cloud motion.

\begin{figure}
\centering
\includegraphics[scale=.8]{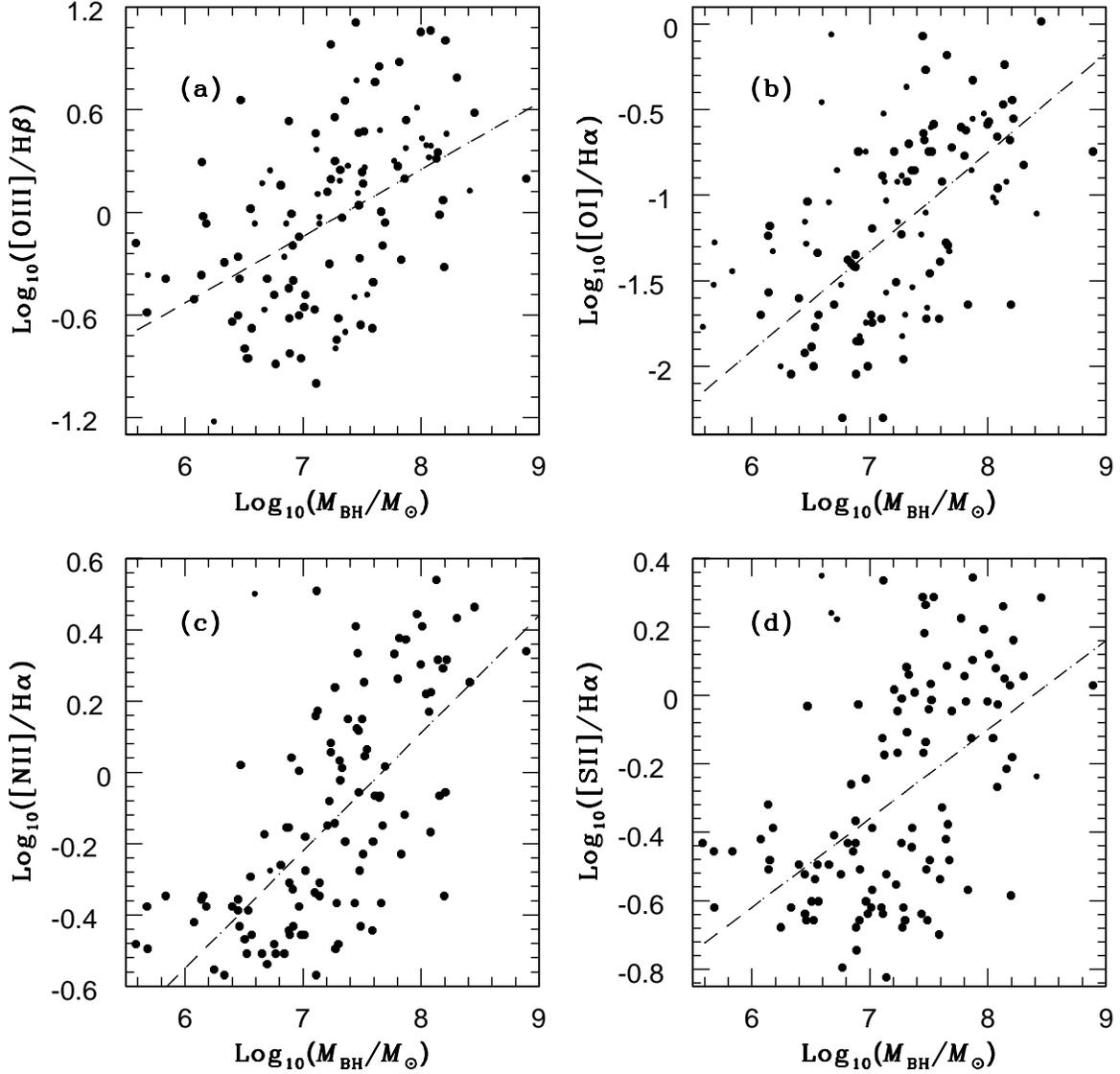}
\caption{BH mass vs.~emission-line ratios. Dashed lines are the linear
least-squares fitting results. Galaxies whose emission-line
ratios are upper or lower limits, or with uncertainties
larger than 30\%, are shown as smaller dots. (a) BH mass
vs.~[O\,{\small III}]/H$\beta$, (b) BH mass vs.~[O\,{\small
I}]/H$\alpha$, (c) BH mass vs.~[N\,{\small II}]/H$\alpha$, and (d)
BH mass vs.~[S\,{\small II}]/H$\alpha$. \label{BH-em}}
\end{figure}

The reason for the unanticipated good correlations between
emission-line ratios and BH mass shown in Figure \ref{BH-em} could
be because the shape of the ionizing continuum is changing, i.e., is
growing harder, with BH mass, but this is only speculation.

The far-infrared (FIR) color indices do not appear well correlated
with BH mass, though the FIR luminosity does.  This could be because
the far-infrared emission likely results from dust heated either by
nuclear star formation or by the active nucleus itself. More
information is required to decide between the two mechanisms.

\begin{figure}
\centering
\includegraphics[scale=.8]{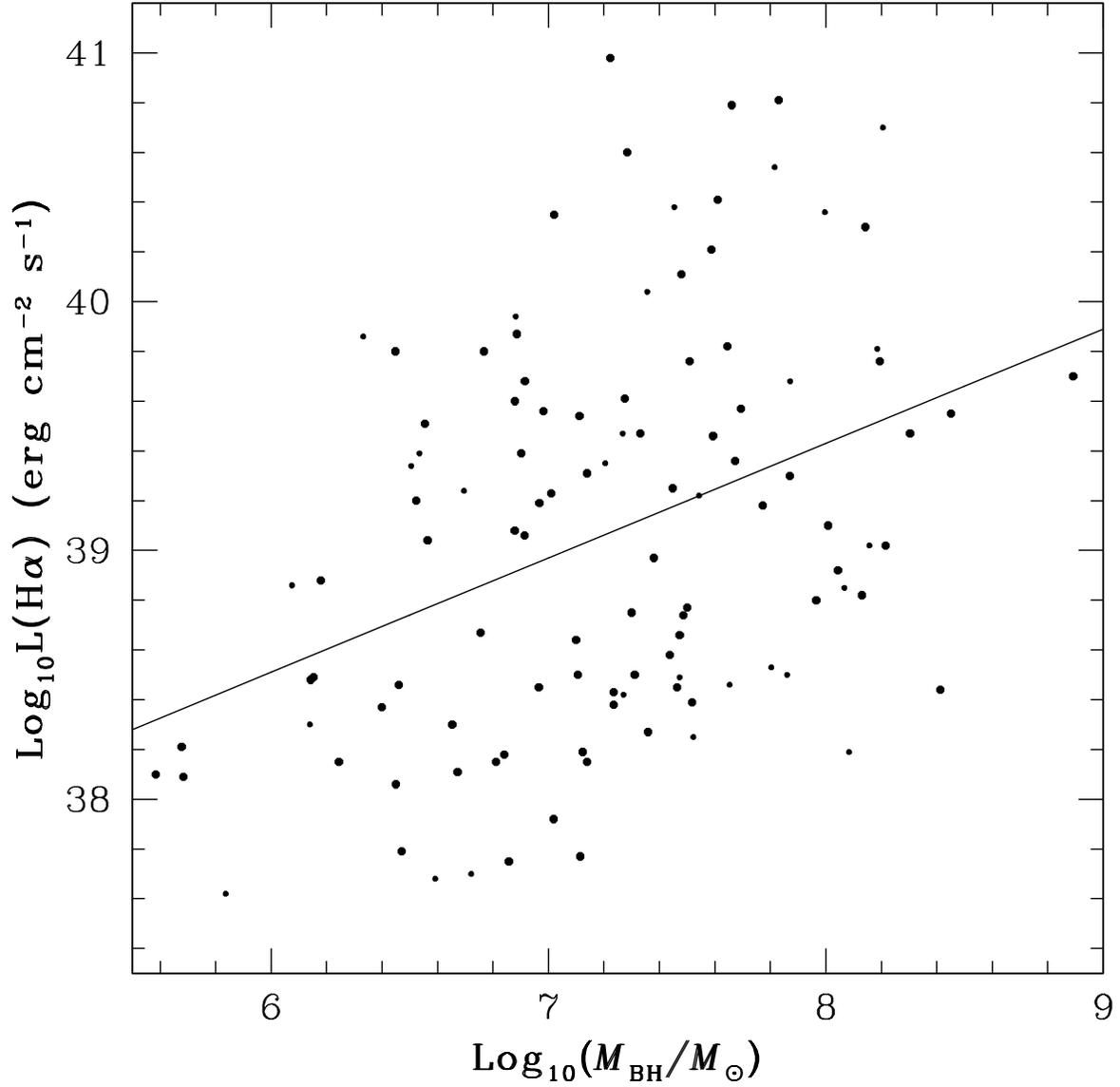}
\caption{BH mass vs.~narrow-line H$\alpha$ luminosity. The solid
line is the linear least-squares fitting result. Galaxies whose
H$\alpha$ luminosities are upper or lower limits, or with
uncertainties larger than 30\%, are shown as smaller
dots.\label{BH-Halpha}}
\end{figure}

More information is also needed to explain the good correlation
shown in Figure \ref{BH-Halpha} between the narrow $\rm{H\alpha}$
emission-line luminosity and the BH mass correlation because narrow
$\rm{H\alpha}$ emission could arise in gas photoionized by the
nucleus and/or from nuclear star formation.

\begin{figure}
\centering
\includegraphics[scale=.8]{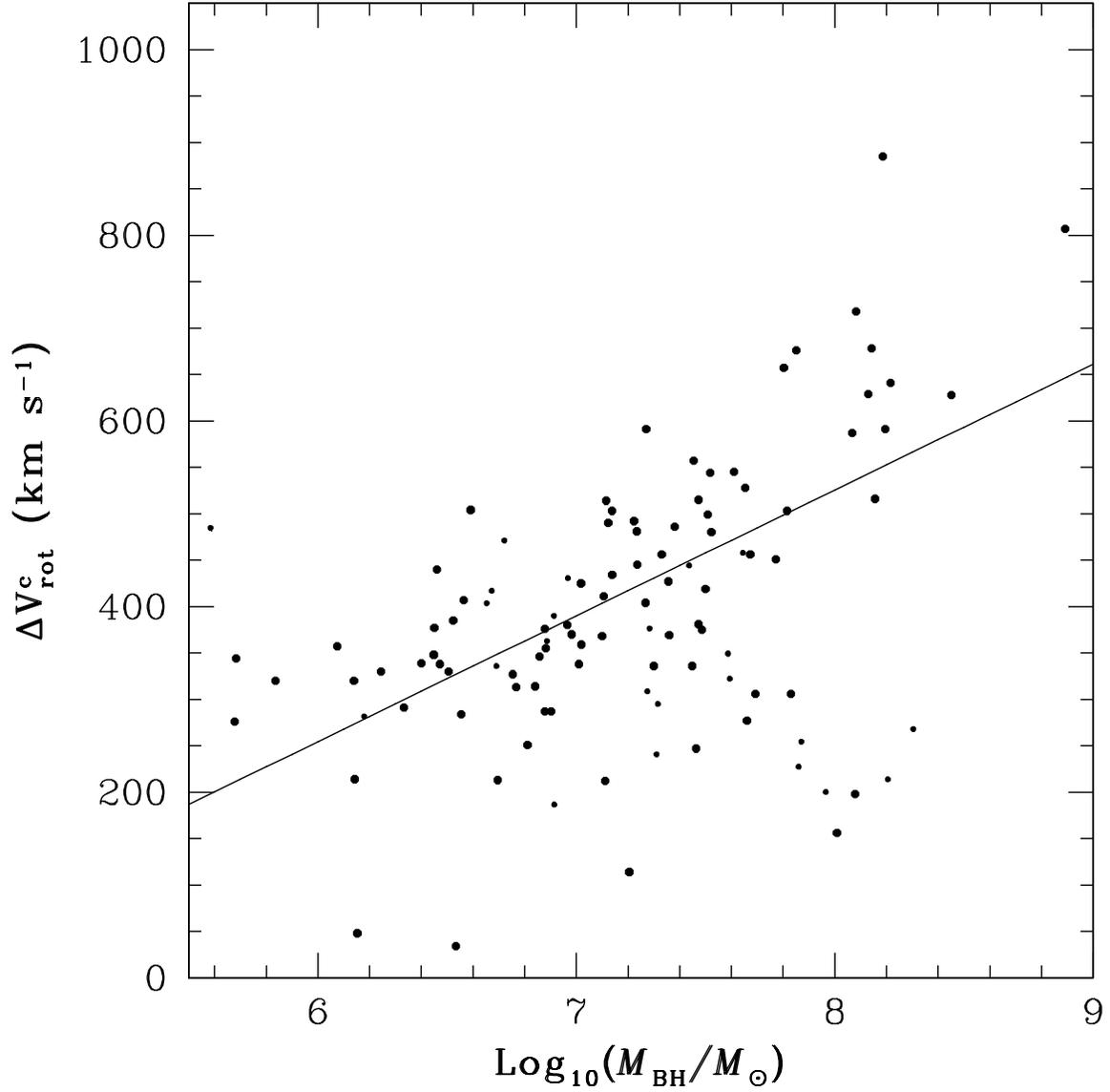}
\caption{BH mass vs.~inclination-corrected H I velocity. The solid
line is the linear least-squares fitting result. Galaxies whose
inclination-corrected H I velocities are lower limits
or with large uncertainties, are shown as smaller dots. \label{BH-dV}}
\end{figure}

The BH mass and the inclination-corrected rotation velocity $\Delta
V^c_{rot}$ are also strongly correlated, with $P<10^{-4}$. This is
not surprising since $\Delta V^c_{rot}$ is a measure of the total
mass within the H\,{\small I} radius of its host galaxy.  This is
shown in Figure \ref{BH-dV}.  Because the rotation velocity normally
is measured at a radius much greater than the bulge scale length,
this correlation might point to relations among the mass of the dark
matter halo, the mass of the bulge, and the mass of the central
black hole.  A correlation of this kind was first noticed by \citeauthor{Ferrarese2002} \citeyearpar{Ferrarese2002}.

No other significant correlations were found between the S\'{e}rsic
index $n$ and other structure parameters (i.e., $R_e$ and $R_s$).
The reason for this is likely because the S\'{e}rsic index was not
always a free parameter in our analysis.

\subsection{Distributions}

The distributions of various parameters as a function of Hubble
stage among the three samples, A, B, and C, were compared using the
non-parametric Kolmogorov-Smirnov (K-S) test \citep{Press1997}. In
the following, $P$ is the probability that two unrelated
distributions would be this similar by chance. A small value of $P$
indicates there is a significant difference between two
distributions.  Data reported as upper or lower limits were not used
in our calculations. Table \ref{distributions} shows the results of
the K-S test among three samples. (A, B) is the probability between
the distributions of sample A and sample B; (A, C) is the
probability between sample A and sample C; and (B, C) is the
probability between sample B and sample C.

\begin{figure}
\centering
\includegraphics[scale=.6]{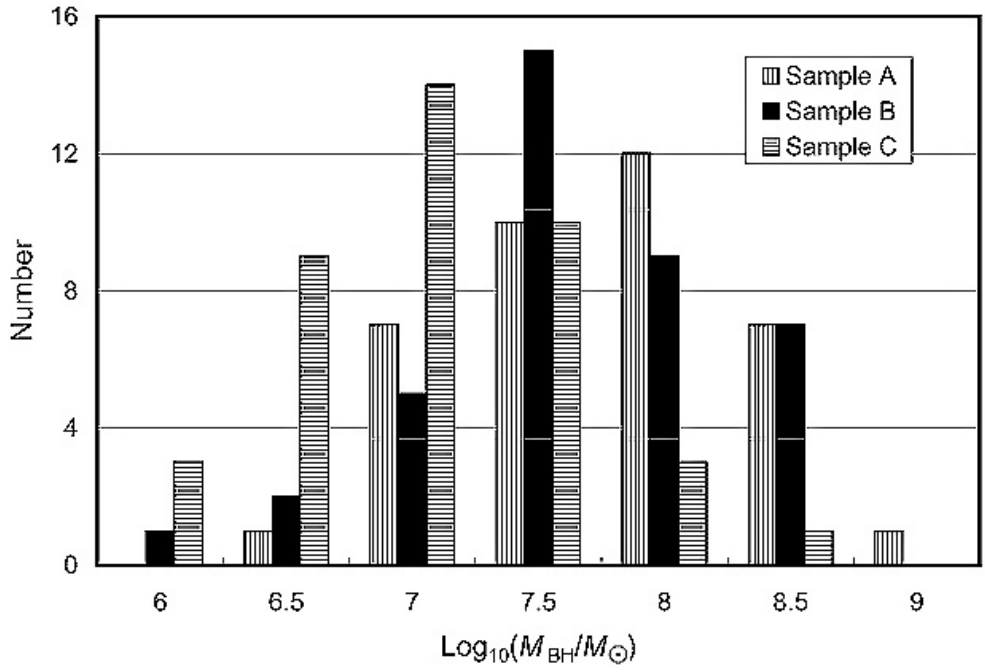}
\caption{Histogram of the BH mass distribution. The x axis shows the
BH mass, while the y axis shows the number of galaxies at this mass.
\label{BH-T-H}}
\end{figure}

The $M_{\rm BH}$, M$_{bul}$ distributions show a strong difference
between samples A and C, and B and C in Figure \ref{BH-T-H}. This
almost certainly means that the bulges (and therefore black holes)
are more massive in sample A and B galaxies, than sample C. This
likely also explains why AGNs are found preferentially in early type
galaxies as the following calculation illustrates.

Using data from local clusters, \citeauthor{deLapparent2003}
\citeyearpar{deLapparent2003} and \citeauthor{Sandage1985}
\citeyearpar{Sandage1985} have quantified the luminosity functions
of early-type and late-type spiral galaxies. In particular, they
suggest that the differential luminosity function, $\phi({\rm M})$,
is Gaussian with $\phi(\rm M) \propto \phi_o \ \exp{(-\frac{({\rm
M-M_o})^2}{\Sigma_o^2})}$, where $\phi_{\rm o}$ is the space density
of a particularly morphological type of spiral galaxy, M$_{\rm o}$
is the mean absolute magnitude, and $\Sigma_{\rm o}$ is the
dispersion of the luminosity function in magnitudes.  In clusters
such as Virgo, Centaurus and Fornax, the space densities of Sa+Sb
and Sc galaxies are similar, while their average absolute magnitudes
differ in the sense that Sa+Sb galaxies are brighter than Sc
galaxies by 1.4 magnitudes in optical bands.

The average absolute $K_s$ magnitudes of galaxies in our samples A,
B and C, however, are similar to within uncertainties: $-23.6 \pm
0.9$, $-23.8 \pm 0.7$ and $-23.3 \pm 0.7$. The average distances of
each sample are also identical to within uncertainties.  The average
bulge luminosities {\it are} different, however, among the samples:
$-22.3 \pm 1.0$ for sample A; $-21.9 \pm 1.0$ for sample B, and
$-20.8 \pm 1.2$ for sample C.  While our sample was not intended to
be complete in any sense, and it should be recalled that 15\% of
Sample C galaxies were not included in this study because they
lacked a perceptible bulge, the sample is primarily representative
of the {\it field} rather than a cluster environment.

We can estimate the relative number of Sa+Sb galaxies compared with
Sc galaxies under the assumption that their luminosity functions and
parameters are as described above.  Using Equation(\ref{Ks-BH}), for
a given BH mass, the absolute $K_s$ magnitude of the bulge can be
computed. The ratio of the space densities of Sa+Sb to Sc galaxies
with bulges as bright or brighter (or equivalently, black hole
masses as great or greater) than this can then be calculated.  Table
\ref{lumin-fun} shows the results of this exercise.

It is clear in a statistical sense that spiral galaxies with central
BH masses in the range of classical AGNs (i.e., $> 10^7~M_\odot$)
occur overwhelmingly in early-type systems. Comparing these data
with the relative number of classical AGNs is problematic; one would
have to understand how accretion rates depend on morphological type,
and make allowances for the heterogeneous ways these systems are
discovered.

\begin{figure}
\centering
\includegraphics[scale=.8]{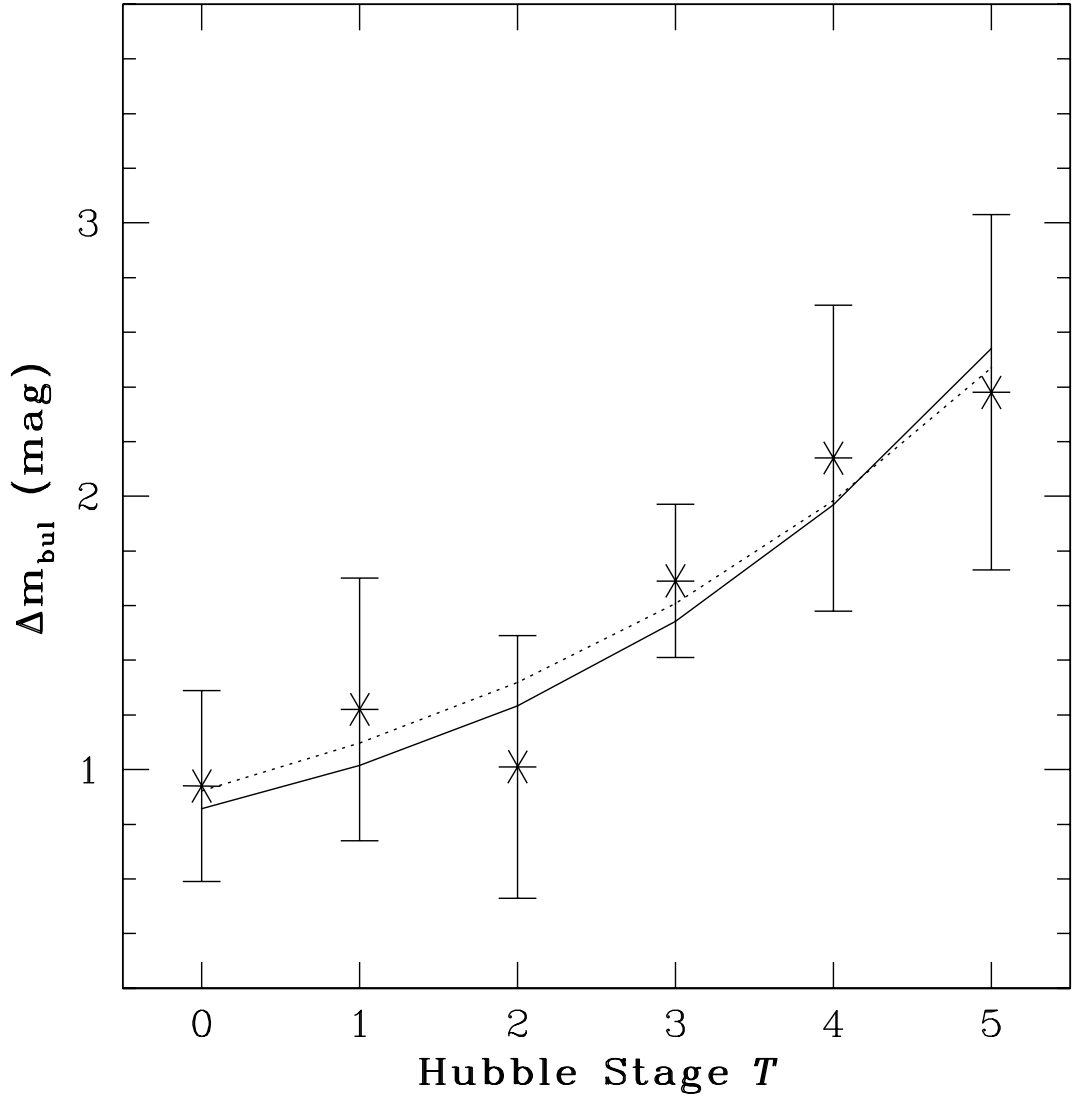}
\caption{$\Delta$m$_{bul}$ vs.~The Hubble Stage $T$ for the $K_{\rm s}$
data showing 1-$\sigma$
standard deviations. The solid line is the best-fit cubic function to the Simien \& de Vaucouleurs' relation in the $B$ band.  
The dashed line is the best-fit cubic function for the $K_{\rm s}$-band
relation described in the text. \label{dm-B-K}}

\end{figure}

A comparison was also made between $\Delta m_{bul}$ given by Simien
\& de Vaucouleurs' empirical relationship \citep{Simien1986} from
Equation (\ref{dmbul}) and (\ref{mbul}) where $m_{bul}$ are from the
GALFIT measurements herein. The original Simien \& de~Vaucouleurs'
relation was measured in the $B$ band, while our sample allows us to
measure the new Simien \& de~Vaucouleurs' relation in the $K_s$
band. Following Simien \& de~Vaucouleurs' formalism, the best-fit
$K_s$ relation is shown in the following cubic equation, where
\begin{equation}
\Delta m_{bul}=0.297(T+5)-0.040(T+5)^2+0.0035(T+5)^3.\end{equation}
The results are illustrated in Figure \ref{dm-B-K}; the solid line
is taken from the Simien \& de~Vaucouleurs' relation in the $B$ band
\citep{Simien1986}, the dashed line is the new relation in $K_s$
band, while our empirical data are illustrated with 1-$\sigma$
standard deviations. It can be seen that there is excellent
agreement between the $B$ and $K_s$-band data and that there is a
very large scatter at each morphological stage.

The FIR luminosities show significantly different distributions
among samples A and B\&C.  FIR emission is normally considered to
originate from dust heated by stars and/or an active nucleus. Sample
A and B have more activity in their nuclei, while sample C galaxies
have more gas and star formation (most of them being H\,{\small II}
region galaxies). Sample B galaxies perhaps show both mechanisms are
present.

\begin{figure}
\centering
\includegraphics[scale=.8]{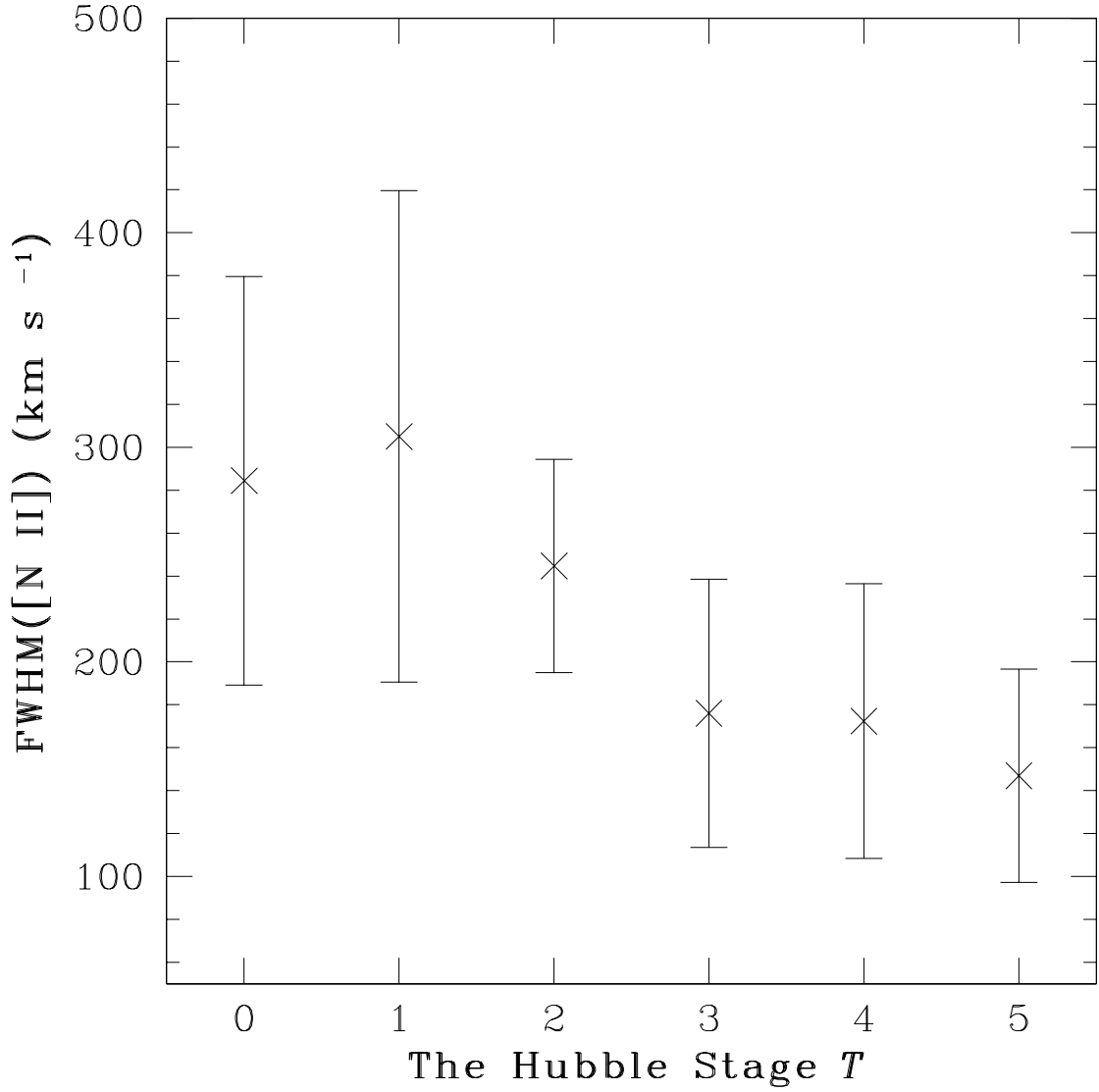}
\caption{FWHM([N\,{\small II}]) distribution along the Hubble stage
$T$ showing 1-$\sigma$ standard deviations. \label{FWHM-T}}
\end{figure}

The FWHM([N\,{\small II}]) distributions are different in all three
populations. In Figure \ref{FWHM-T} we show the mean
FWHM([N\,{\small II}]) for each Hubble stage and the accompanying
standard deviation. In early-type galaxies with $T<3$, BHs may
contribute to the line width, i.e., the more massive the BH, the
broader the random velocities and hence emission lines. For
later-type galaxies with $T\geq 3$, the FWHM appears almost
constant, possibly reflecting the rotational and random velocities
of the bulges alone. The [O\,{\small III}]/H$\beta$ and [N\,{\small
II}]/H$\alpha$ ratios show clear differences between samples A\&B
and C galaxies. These ratios indicate the ionization level in the
emission-line gas and so the relative contribution of nonthermal
vs.~thermal radiation.

The inclination-corrected rotation velocities are different in
samples A\&B and C. The reason is likely because samples A and B
galaxies have more massive bulges than C galaxies.

The G-band distributions show that sample A\&B are similar, while
sample C is different. The G-band begins to become significant in
early or mid-F spectral type stars, and becomes prominent in late-F
to K stars.  It is weak in early-type stars and star-forming
regions. The difference among the distributions, although not
strong, could provide an indication that the integrated populations
are different in these subsamples.  Sample C galaxies generally have
a weaker G-band that is indicative of a younger population.

As expected, there are no statistically significant differences
involving galaxy density $\rho_{gal}$, the projected angular
separation between the galaxy and its neighbor $\theta_p$, and the
inclination of the disk with distance.

We also investigated the role played by the level of nuclear
activity among the various parameters.  Of 118 galaxies, 19 are
classified as Seyferts, 26 as LINERs, 57 as H\,{\small II} region
galaxies, and 15 as transition objects (HFS).  In most correlations
and distributions, H\,{\small II} region galaxies can be
clearly distinguished from the other galaxies, but there is no
significant grouping among the other (AGN) categories.

\section{Summary and Conclusions}

Through the calibration of the black hole --- $K_s$ bulge luminosity
relation, we determined the central black hole masses for 118 spiral
galaxies using 2MASS data for a variety of morphological stages from
the spectroscopic survey of HFS.  The bulge luminosities were
measured using GALFIT, a two-dimensional bulge/disk decomposition
routine.

Nuclear properties such as line width (FWHM([N\,{\small II}])), as
well as emission-line ratios (e.g., [O\,{\small III}]/H$\beta$,
[O\,{\small I}]/H$\alpha$, [N\,{\small II}]/H$\alpha$, and
[S\,{\small II}]/H$\alpha$), showed a very high degree of
correlation with black hole mass.  The excellent line-width
correlation provides strong support that the emission-line gas is in
virial equilibrium with either the black hole or bulge potential.
The very good emission-line ratio correlations seem to suggest that
more massive black holes give rise to harder ionizing radiation.

The only non-trivial host-galaxy parameter that correlated well with
black-hole mass is the inclination-corrected 
rotational velocity.  This may suggest that the black hole-bulge 
relation may also extend to the dark matter halo.

The sample was divided into three subsamples, A, B, and C according
to host-galaxy morphology.  Significant differences were found among
the distributions of black hole masses in the subsamples in the
sense that early-type galaxies have preferentially larger central
black holes.  This is consistent with observations that Seyfert
galaxies, for example, are found preferentially in early-type
systems.  The line-width distributions also illustrated clear
differences among subsamples A, B and C, in the sense that
earlier-type galaxies have larger line widths, widths that could
have a significant contribution by the central black hole, as well
as the bulge potential.

Marked differences were found between subsamples A+B and C for
emission-line ratios, differences that could be attributed to the
level of ionization in the gas in the sense that late-type galaxies
have a larger thermal component.  Finally, a $K_s$-band Simien \& de
Vaucouleurs diagram showed excellent agreement with the original
$B$-band relation, with a considerable dispersion at every
morphological stage.

The authors would like to thank the Natural Sciences and Engineering
Research Council of Canada for support, as well as C.J. Ryan for
insightful discussions.  This publication makes use of data products
from the Two Micron All Sky Survey, which is a joint project of the
University of Massachusetts and the Infrared Processing and Analysis
Center/California Institute of Technology, funded by the National
Aeronautics and Space Administration and the National Science
Foundation.  This work was based on research carried out by X.Y.D. in
partial fulfillment of an M.Sc. thesis.

\clearpage

\clearpage

\clearpage

\begin{deluxetable}{ccccccccc}
\tabletypesize{\scriptsize} \tablewidth{0pc}
\tablecaption{Emission-line and host-galaxy parameters
\label{em-host}} \tablehead{ \colhead{Galaxy Name} &
\colhead{Log$L$(H$\alpha$)} & \colhead{FWHM([N\,{\scalebox{.8}
{II}}])} & \colhead{H$\alpha$/H$\beta$} &
\colhead{[O\,{\scalebox{.8} {III}}]/H$\beta$} &
\colhead{[N\,{\scalebox{.8} {II}}]/H$\alpha$}  &
\colhead{$T$} & \colhead{$d$} & \colhead{$\Delta V^c_{rot}$}\\
\colhead{} & \colhead{} & \colhead{(km s$^{-1}$)} & \colhead{} &
\colhead{} & \colhead{} & \colhead{} & \colhead{(Mpc)} &
\colhead{($\rm{km s^{-1}}$)}\\ \colhead{(1)} & \colhead{(2)} &
\colhead{(3)} & \colhead{(4)} & \colhead{(5)} & \colhead{(6)} &
\colhead{(7)} & \colhead{(8)} & \colhead{(9)}} \startdata
&&&\textbf{Sample A}&&&&&\\
\\
NGC 1358    &   40.36L  &   246 &   3.72    &   11.33   &   2.01    &   0   &   53.6    &   \nodata \\
NGC 2639    &   39.68L  &   401 &   4.06    &   3.46    &   4.25    &   1   &   42.6    &   \nodata \\
NGC 2655    &   39.55   &   404 &   4.94    &   3.83    &   2.91    &   0   &   24.4    &   628 \\
NGC 2782    &   40.79   &   312 &   6.5 &   1.01    &   0.43    &   1   &   37.3    &   277 \\
NGC 3166    &   39.1    &   344 &   6.50b   &   2.71b   &   2.57    &   0   &   22  &   156 \\
NGC 3169    &   39.02   &   348 &   5.03b   &   2.88b   &   2.07    &   1   &   19.7    &   641 \\
\enddata
\tablecomments{Emission-line and host-galaxy parameters were taken
from HFS. H$\alpha$ and H$\beta$ refer to the narrow emission-line
components.  Col.(1): Galaxy name. Col.(2):
Luminosity of the narrow component of H$\alpha$ in units of ${\rm
ergs~s^{-1}}$. Col.(3): FWHM of
[N\,{\scalebox{0.8}{II}}]$\lambda$6583. Col.(4): H$\alpha$/H$\beta$,
corrected for Galactic reddening. Col.(5),(6): Dereddened (Galactic
and internal) emission-line intensity ratios. The symbols have the
following meanings: `L'$=3\sigma$ lower limit; `u'$=3\sigma$ upper
limit. Values with uncertainties of $\pm30\%-50\%$ are followed by
`b', while highly uncertain values with probable errors of
$\pm$100\% are followed by `c'. Col.(7): Hubble morphological stage
index. Col.(8): Distance. Col.(9): Rotational velocity amplitude corrected
for inclination: `a' refers to a lower limit because the
inclination is not available; `b' suggests the
inclination correction uncertain and is set to 30$^{\circ}$;
`c' means possibly confused by another nearby galaxy;
`d' indicates the velocity width was measured at 50\% of the
peak of the H\,{\scalebox{0.8}{I}} profile.}
\end{deluxetable}
\clearpage

\begin{deluxetable}{lcccc}
\tabletypesize{\scriptsize} \tablecaption{List of BHs used to
calibrate the BH mass and the bulge magnitude correlation
\label{BH-calibration}} \tablewidth{0pt}
\tablehead{ \colhead{Galaxy} & \colhead{Distance} & \colhead{$M_{\rm BH}$} & \colhead{Reference} & \colhead{${\rm M}_{bul}$}\\
\colhead{} & \colhead{(Mpc)} & \colhead{($10^{8}M_{\rm \odot}$)} &
\colhead{} & \colhead{(mag)}\\
\colhead{(1)} & \colhead{(2)} & \colhead{(3)} & \colhead{(4)} &
\colhead{(5)}} \startdata

NGC 5845   &   24.74   &   2.4 $_{-1.4}^{+0.4}$    &   1   &   -22.86 $\pm$   0.21    \\
NGC 221    &   0.77    &   0.025   $_{-0.005}^{+0.005}$    &   2   &   -19.34  $\pm$   0.08    \\
IC 1459   &   28.83   &   26  $_{-11}^{+11}$  &   3   &   -25.49  $\pm$   0.28    \\
NGC 4374   &   17.72   &   17  $_{-6.7}^{+12}$ &   4   &   -25.02  $\pm$   0.11    \\
NGC 4697   &   11.43   &   1.7 $_{-0.3}^{+0.2}$    &   1   &   -23.92  $\pm$   0.14    \\
NGC 4649   &   16.44   &   20  $_{-6}^{+4}$    &   1   &   -25.34  $\pm$   0.15    \\
NGC 4261   &   31.12   &   5.4 $_{-1.2}^{+1.2}$    &   5   &   -25.20  $\pm$   0.19    \\
NGC 4564   &   14.53   &   0.56    $_{-0.08}^{+0.03}$  &   1   &   -22.88  $\pm$   0.17    \\
\enddata
\end{deluxetable}

\begin{deluxetable}{lcccccccc}
\tabletypesize{\scriptsize} \tablecaption{GALFIT Results
\label{GALFIT-result}} \tablewidth{0pt} \tablehead{ \colhead{Galaxy
Name} & \colhead{Func.} & \colhead{$m_{bul}$ or $m_{disk}$}  &
\colhead{$R_e$ or $R_s$}  & \colhead{$n$}   & \colhead{$\chi^2$}
&   \colhead{$m_{2MASS}$}  &   \colhead{$m_{mod}$}  &   \colhead{residual flux}\\
\colhead{}&\colhead{}&\colhead{}&\colhead{(arcsec)}&\colhead{}&\colhead{}&\colhead{}&\colhead{}&\colhead{(percentage)}\\
\colhead{(1)} & \colhead{(2)} & \colhead{(3)} & \colhead{(4)} &
\colhead{(5)} & \colhead{(6)} & \colhead{(7)} & \colhead{(8)} &
\colhead{(9)}}
\startdata &&&&\textbf{Sample A}&&&&\\
\\
NGC 1358    &   Sersic  &   10.18   &   4.51    &   2.28    &   0.01    &   8.948   &   9.09    &   14.3    \\
\nodata &   exp &   9.59    &   21.20   &   \nodata &   \nodata &   \nodata &   \nodata &   \nodata \\
NGC 2273    &   Sersic  &   9.66    &   3.60    &   5.50    &   0.04    &   8.480   &   8.47    &   -1.0    \\
\nodata &   exp &   8.91    &   15.61   &   \nodata &   \nodata &   \nodata &   \nodata &   \nodata \\
NGC 2639    &   Sersic  &   9.96    &   8.21    &   4.00    &   1.26    &   8.397   &   8.44    &   4.2     \\
\nodata &   exp &   8.75    &   10.02   &   \nodata &   \nodata &   \nodata &   \nodata &   \nodata \\
\enddata

\tablecomments{Col.(1): Galaxy name. Col.(2): Profile function.
Col.(3): Bulge magnitude with the average formal error about 1.4\%
of the disk magnitude with the average formal error about 0.7\%.
Col.(4): $R_e$, half-light effective radius for the bulge component,
with the average formal error about 9.3\% ; $R_s$, scale length for
the disk component, with the average formal error about 2\%; in
units of arcsec. Col.(5): S\'{e}rsic index.
Col.(6): $\chi^{2}_\nu$. Col.(7): total
magnitude from 2MASS, with the average error about 0.027 mag.
Col.(8): total magnitude of the model. Col.(9): Residual flux,
\begin{math}r=10^{-0.4(m_{2MASS}-m_{mod})}-1\end{math}.}

\end{deluxetable}
\clearpage

\begin{deluxetable}{ccccccc}
\tabletypesize{\scriptsize} \tablecaption{The Estimated BH Mass
\label{BH-mass}} \tablewidth{0pt} \tablehead{ \colhead{Galaxy Name}
& \colhead{${\rm M}_{tot}$}    & \colhead{${\rm M}_{bul}$}  &
\colhead{$R_e$}  & \colhead{${\rm M}_{disk}$} &   \colhead{$R_s$}  &
\colhead{${\rm Log_{10}}
(\it M_{\rm BH}/M_\odot)$}\\
\colhead{}    &  \colhead{(mag)}     &  \colhead{(mag)}     &
\colhead{(kpc)}    &  \colhead{(mag)}     & \colhead{(kpc)}    &
\colhead{}\\ \colhead{(1)} & \colhead{(2)} & \colhead{(3)} &
\colhead{(4)} & \colhead{(5)} & \colhead{(6)} & \colhead{(7)}} \startdata &&&\textbf{Sample A}&&&\\
\\
NGC 1358    &   -24.6   &   -23.3   &   1.17    &   -23.9   &   5.51    &   8.0     \\
NGC 2273    &   -23.6   &   -22.5   &   0.50    &   -23.2   &   2.15    &   7.6     \\
NGC 2639    &   -24.6   &   -23.0   &   1.70    &   -24.3   &   2.07    &   7.9     \\
NGC 2655    &   -24.8   &   -24.3   &   2.34    &   -23.8   &   5.06    &   8.5     \\
NGC 2782    &   -23.8   &   -22.6   &   0.63    &   -23.5   &   2.76    &   7.7     \\
NGC 3166    &   -24.4   &   -23.4   &   0.64    &   -23.4   &   4.24    &   8.0     \\
\enddata

\tablecomments{Col.(1): Galaxy name. Col.(2): Absolute magnitude in
$K_s$ band for the total galaxy. Col.(3): Absolute magnitude in
$K_s$ band for the galaxy bulge. Col.(4): Half-light radius of the
bulge in units of kpc. Col.(5): Absolute magnitude in $K_s$ band for
the galaxy disk. Col.(6): Scale length of the disk in units of kpc.
Col.(7): Estimated BH mass.}

\end{deluxetable}
\clearpage

\begin{deluxetable}{llcccc}
\tabletypesize{\scriptsize} \tablecaption{Correlations between
parameters \label{correlations}} \tablewidth{0pt} \tablehead{
\colhead{X} & \colhead{Y} & \colhead{SLOPE} & \colhead{INTERCEPT} & \colhead{r} & \colhead{\it P}\\
\colhead{} & \colhead{} & \colhead{(uncertainty)} &
\colhead{(uncertainty)} & \colhead{} & \colhead{}} \startdata
Log$_{10}$(\it M$_{\rm BH}$/M$_{\odot}$)& \rm Log$_{10}$(FWHM([N\,{\scalebox{.8} {II}}]) & 0.24($\pm$0.03) & 0.50($\pm$0.19) & 0.69 & 9.4$\times10^{-16}$\\
Log$_{10}$(\it M$_{\rm BH}$/M$_{\odot}$)& Log$_{10}$([O\,{\scalebox{.8} {III}}]/H$\beta$) & 0.39($\pm$0.08) & -2.9($\pm$0.58) & 0.49 & 4.8$\times10^{-6}$\\
Log$_{10}$(\it M$_{\rm BH}$/M$_{\odot}$)& Log$_{10}$([O\,{\scalebox{.8} {I}}]/H$\alpha$) & 0.58($\pm$0.09) & -5.4($\pm$0.63) & 0.63 & 4.1$\times10^{-9}$\\
Log$_{10}$(\it M$_{\rm BH}$/M$_{\odot}$)& Log$_{10}$([S\,{\scalebox{.8} {II}}]/H$\alpha$) & 0.26($\pm$0.04) & -2.2($\pm$0.29) & 0.55 & 8.5$\times10^{-10}$\\
Log$_{10}$(\it M$_{\rm BH}$/M$_{\odot}$)& Log$_{10}$([N\,{\scalebox{.8} {II}}]/H$\alpha$) & 0.33($\pm$0.03) & -2.5($\pm$0.25) & 0.69 & 9.6$\times10^{-17}$\\
Log$_{10}$(\it M$_{\rm BH}$/M$_{\odot}$)& Log$_{10}$$L(S_{12})$ & 0.30($\pm$0.07) & 0.20($\pm$0.48) & 0.44 & 1.5$\times10^{-5}$\\
Log$_{10}$(\it M$_{\rm BH}$/M$_{\odot}$)& Log$_{10}$$L(S_{25})$ & 0.34($\pm$0.08) & 0.01($\pm$0.55) & 0.43 & 2.0$\times10^{-5}$\\
Log$_{10}$(\it M$_{\rm BH}$/M$_{\odot}$)& Log$_{10}$$L(S_{60})$ & 0.23($\pm$0.08) & 1.6($\pm$0.56) & 0.28 & 3.6$\times10^{-3}$\\
\enddata

\tablecomments{Col.(1): X dataset when calculating the correlations.
Col.(2): Y dataset when calculating the correlations. Col.(3): Slope
for the linear least-squares fitting. Col.(4): Intercept for the
linear least-squares fitting. Col.(5): Correlation coefficient.
Col.(6): Significance level or probability.}

\end{deluxetable}

\clearpage

\begin{deluxetable}{lccc}
\tabletypesize{\scriptsize} \tablecaption{Parameter Distributions
along the Hubble Stage \label{distributions}} \tablewidth{0pt}
\tablehead{ \colhead{} & \colhead{(A,B)} & \colhead{(A,C)}
&\colhead{(B,C)}} \startdata

\it M$_{\rm BH}$/M$_{\odot}$    & \rm  0.42  &   $1.8\times10^{-5}$    &   $5.9\times10^{-4}$    \\
M$_{bul}$   &   0.29  &   $3.5\times10^{-5}$    &   $6.8\times10^{-4}$    \\
M$_{disk}$  &   0.63  &   0.88  &   0.47  \\
M$_{tot}$   &   0.42  &   0.05  &   0.02  \\
$\Delta V^c_{rot}$  &   0.15  &   0.07  &   0.02  \\
Log$_{10}$$L(\rm{H\alpha})$   &   0.60  &   0.05  &   0.17  \\
Log$_{10}$$L(S_{12})$ &   $4.7\times10^{-3}$    &   $1.2\times10^{-3}$    &   0.91  \\
Log$_{10}$$L(S_{25})$ &   0.06  &   0.01  &   0.81  \\
\enddata

\tablecomments{Col.(1): Parameter name. Col.(2): Significance of
the difference between the distributions of Sample A and Sample B.
Col.(3): Significance of the difference between the distributions of
Sample A and Sample C. Col.(4): Significance of the difference
between the distributions of Sample B and Sample C. If
$P < 0.05$, the two distributions are significantly different.}

\end{deluxetable}

\begin{deluxetable}{ccr}
\tablecaption{AGN Model Luminosity Function Ratio \label{lumin-fun}}
\tablewidth{0pt} \tablehead{ \colhead{Log$_{10}(M_{\rm
BH}/M_{\odot}$)} & \colhead{M$_{bul}$} &
\colhead{$\phi$(Sa+Sb)/$\phi$(Sc)}} \startdata

6.00 & -18.87 &  1.01 \\
6.25 & -19.43 &  1.06 \\
6.50 & -19.98 &  1.20 \\
6.75 & -20.54 &  1.58 \\
7.00 & -21.09 &  2.52 \\
7.25 & -21.65 &  4.63 \\
7.50 & -22.20 &  8.97 \\
7.75 & -22.76 & 16.69 \\
8.00 & -23.31 & 28.12 \\

\tablecomments{Col.(1): Central BH mass. Col. (2): Absolute $K_s$
bulge magnitude. Col (3): Ratio of the space densities of Sa+Sb
galaxies to Sc galaxies with bulges as bright or brighter, or
equivalently, with black hole masses as large or larger, than
listed.}

\enddata
\end{deluxetable}


\clearpage

\begin{thebibliography}{}
\bibitem[Barth et al.(1999)]{Barth1999} Barth, A. J.,
Filippenko, A. V., \& Moran, E. C. 1999, ApJ, 525, 673
\bibitem[Bendo \& Joseph 2004]{Bendo2004} Bendo, G. J. \& Joseph, R. D. 2004, \aj, 127, 3338
\bibitem[Blandford \& McKee 1982]{Blandford1982} Blandford, R. D. \& McKee, C. F. 1982, \apj, 255, 419
\bibitem[Boroson 2003]{Boroson2003} Boroson, T. A. 2003, \apj, 585,
681
\bibitem[Bower et al.(1998)]{Bower1998} Bower, G. et al. 1998, \apj, 492, L111
\bibitem[Cappellari et al.(2002)]{Cappellari2002} Cappellari, M. et al. 2002, \apj, 578, 787
\bibitem[de Vaucouleurs et al. 1976]{deVaucouleurs1976} de Vaucouleurs, G., de Vaucouleurs, A., \& Corwin, J. R. 1976, Second Reference Catalogue of Bright Galaxies. (Austin: The Univ. of Texas
Press.)
\bibitem[de Lapparent(2003)]{deLapparent2003} de Lapprent, V. 2003, A\&A, 408, 845.
\bibitem[Erwin et al. 2004]{Erwin2004} Erwin, P., Graham, A.
W., \& Caon, N. 2004, in Coevolution of Black Holes \& Galaxies, ed.
L. C. Ho (Cambridge: Cambridge Univ. Press), 12
\bibitem[Ferrarese et al.(1996)]{Ferrarese1996} Ferrarese, L., Ford, H. C., \& Jaffe, W. 1996, \apj, 470, 444
\bibitem[Ferrarese \& Ford (1999)]{Ferrarese1999} Ferrarese, L. \& Ford, H. C. 1999, \apj, 515, 583
\bibitem[Ferrarese \& Merritt (2000)]{Ferrarese2000} Ferrarese, L. \& Merritt, D. 2000, \apjl, 539, 9
\bibitem[Ferrarese(2002)]{Ferrarese2002} Ferrarese, L. 2002, \apj, 578, 90
\bibitem[Ferrarese \& Ford (2005)]{Ferrarese2005} Ferrarese, L. \& Ford, H. 2005, Space Science Reviews, 116, 523
\bibitem[Freeman 1970]{Freeman1970} Freeman, K. C. 1970, \apj, 160, 811
\bibitem[Gebhardt et al.(2000)]{Gebhardt2000} Gebhardt, K., et al. 2000, \aj, 119, 1157
\bibitem[Gebhardt et al.(2003)]{Gebhardt2003} Gebhardt, K., et al. 2003, \apj, 583, 92
\bibitem[Ghez 2004]{Ghez2004} Ghez, A. M. 2004, in Coevolution of Black Holes \& Galaxies, ed. L. C. Ho (Cambridge: Cambridge Univ.
Press), 53
\bibitem[Graham 2001]{Graham2001} Graham, A. W. 2001, \aj, 121, 820
\bibitem[Greenhill et al. 1995]{Greenhill1995} Greenhill, L. J., et al. 1995, \apj, 440, 619
\bibitem[Ho et al.(1997)Ho, Fillippenko, and Sargent]{Ho1997} Ho, L. C., Fillippenko, A. V., \& Sargent, W. L. W.  1997, \apjs, 112, 315
\bibitem[Jarrett et al. 2000]{Jarrett2000} Jarrett, T. H., et al. 2000, \aj, 119, 2498
\bibitem[Kaspi et al. 2000]{Kaspi2000} Kaspi, S., et al. 2000, \apj, 533, 631
\bibitem[Kormendy 2004]{Kormendy2004} Kormendy, J. 2004, in Coevolution of Black Holes \& Galaxies, ed.
L. C. Ho (Cambridge: Cambridge Univ. Press), 1
\bibitem[Kormendy \& Gebhardt (2001)]{Kormendy2001} Kormendy, J. \& Gebhardt, K. 2001, AIP Conf. Proc. 586, 20th Texas Symposium on Relativistic Astrophysics, ed. J. C. Wheeler \& H. Martel (New York: AIP), 363
\bibitem[Kormendy \& Richstone (1995)]{Kormendy1995} Kormendy, J. \& Richstone, D. 1995, ARA\&A, 33, 581
\bibitem[Macchetto et al.(1997)]{Macchetto1997} Macchetto, D. F., et al. 1997, \apj, 489, 579
\bibitem[Marconi \& Hunt (2003)]{Marconi2003} Marconi, A. \& Hunt, L. K. 2003, \apjl, 589, 21
\bibitem[Merritt \& Ferraress (2001)]{Merritt2001} Merritt, D. \& Ferrarese, L. 2001, \apj, 547, 140
\bibitem[Moran et al. 1999]{Moran1999} Moran, J. M., Greenhill, L. J., \& Herrnsten, J. R., JApA, 20 165
\bibitem[Nelson \& Whittle 1996]{Nelson1996} Nelson, C. H. \& Whittle, M. 1996, \apj, 465, 96
\bibitem[Peng et al.(2002)]{Peng2002} Peng, C. Y., et al. 2002, \aj, 124, 266
\bibitem[Peterson et al. 2004]{Peterson2004} Peterson, B. J., et al. 2004, \apj, 613, 682
\bibitem[Press et al. 1997]{Press1997} Press, W. H., et
al. 1997, Numerical Recipes in C. (Cambridge: Cambridge Univ.
Press.)
\bibitem[Rees 1984]{Rees1984} Rees, M. J. 1984, \araa, 22, 471
\bibitem[Sandage \& Tammann 1981]{Sandage1981} Sandage, A. \& Tammann, G. A. 1981, {\it A Revised Shapley-Ames Catalog of Bright
    Galaxies.} Washington: Carnegie Institution of Washington.
\bibitem[Sandage et al.(1985)]{Sandage1985} Sandage, A., Binggeli, B., \& Tammann, G.A. 1985, AJ, 90, 1759
\bibitem[Schlegel et al. 1998]{Schlegel1998} Schlegel, D., J., Finkbeiner, D. P., \& Davis, M. 1998, \apj, 500, 525
\bibitem[S\'{e}rsic 1968]{Sersic1968} S\'{e}rsic, J. L. 1968, Atlas de Galaxias Australes (C\'{o}rdoba: Obs. Astron., Univ. Nac.
C\'{o}rdoba.)
\bibitem[Simien \& de Vaucouleurs 1986]{Simien1986} Simien, F. \& de Vaucouleurs, G. 1986, \apj, 302, 564
\bibitem[Tonry et al.(2001)]{Tonry2001} Tonry, J. L., et al. 2001, \apj,
546, 681
\bibitem[Tremaine et al.(2002)]{Tremaine2002} Tremaine, S. et al. 2002, \apj, 574, 740
\bibitem[Tully (1988)]{Tully1988} Tully, R. B. 1988, Nearby Galaxies Catalog (Cambridge: Cambridge Univ.
Press.)
\bibitem[Tully \& Shaya 1984]{Tully1984} Tully, R. B. \& Shaya, E. J. 1984, \apj, 281, 31
\bibitem[Valluri et al. 2004]{Valluri2004} Valluri, M., Merritt, D., \& Emsellem, E. 2004, \apj, 602, 66
\bibitem[van der Marel \& van den Bosch(1998)]{vanderMarel1998} van der Marel, R. P. \& van den Bosch, F. C. 1998, \aj, 116, 2220
\bibitem[Verolme et al.(2002)]{Verolme2002} Verolme, E. K., et al. 2002, MNRAS, 335, 517
\bibitem[Wandel et al. 1999]{Wandel1999} Wandel, A., Peterson, B. M., \& Malkan, M. A. 1999, \apj, 526, 579
\end{thebibliography}
\end{document}